\documentclass[a4paper]{article}
\usepackage[T1]{fontenc}
\usepackage[utf8]{inputenc}
\usepackage{utopia}
\usepackage{eurosym}
\usepackage{url}
\usepackage{typearea}
\usepackage[protrusion=true, expansion=false]{microtype}

\typearea{8}

\title{Juppix:\\a Linux Live-CD for Undergraduate Students}
\author{Juliusz Chroboczek and Sylvain Lebresne\\{\tt juppix@pps.jussieu.fr}}
\date{18 November 2008}

\begin{document}

\maketitle

\begin{abstract}
  Juppix is a Linux Live-CD with a comfortable programming environment for
  the Java, C and O'Caml programming languages that has been distributed to
  hundreds of undergaduate students at the University of Paris 7 over the
  last few years.  We describe the lessons we learnt while compiling and
  distributing Juppix, and outline our future plans.
\end{abstract}

\section{Background and motivation}

Undegraduate teaching of programming at the University of Paris 7 is mostly
based on Unix.  During the first year, we use the Java programming language
with some locally-developed packages \cite{classe-deug} and the Emacs
editor, in the second year, we add C and PHP, and third year students get
to enjoy O'Caml.

While on campus, our students do have access to a few dozen reasonably
configured machines running a system consisting of FreeBSD Unix and the KDE
desktop.  In periods of high affluence, such as just before a project
deadline, our computer rooms tend to be overcrowded.  Even in normal times,
many students prefer to work from the comfort of their homes or to sit with
their laptop in a nearby café.

It used to be the case that well-meaning lecturers and tutors helped
students set up a Linux distribution on their personal machine.
Unfortunately, destroying a student's MP3 collection by mistake is not
a good way of staying popular, and visiting a student's home (especially
a female student's home), even with the laudable goal of helping with
a Linux installation, makes some of us uncomfortable.

Perhaps more importantly, many first year students are overwhelmed with the
change in habits that we require from them: after years of clicking on the
pretty icons of a Microsoft Windows environment, they are suddenly required
to use a fairly sober KDE desktop, the command line and Emacs.  The slight
differences between the carefully tuned configuration that they access at
the university and the default configuration of a general-purpose Linux
distribution are often enough to confuse them.

\paragraph{Live-CDs} A Live-CD is a bootable CD that provides a full
operating system environment that can be run from the CD with no need to
install any software on a computer's hard disk.  While Linux Live-CDs have
been around for as long as anyone can remember, with some very early Linux
distributions using a Live-CD format \cite{yggdrasil-live-cd}, they have
only become popular in recent years, probably because of CD-ROM drives
becoming fast enough.

In recent years, the Live-CD concept has become increasingly popular, and
there is a vast array of available Live-CD implementations of
general-purpose operating systems, but also of appliance software such as
routers, firewalls and kiosks to be deployed in public places.  The concept
has become popular enough for at least one major Linux distribution to risk
using a Live-CD as their primary installation media \cite{ubuntu-live-cd}.

Our department had had some extensive experience with Live-CDs in the late
nineties and early naughts \cite{demolinux}.  It was natural for us to
think of distributing software to our students in this format.

\section{The early days}

In the autumn of 2004, the second author started experimenting with using
a {\em Knoppix\/} Linux LiveCD \cite{knoppix} and manually replacing some
little-used pieces of software (such as the KDE games and network security
tools) with a full set of packages for Java programming.  He distributed
a few copies of these to members of his tutorial groups, encouraging them
to try this at home.

When his CDs turned out to be fairly successful with his students, the
first author convinced him to turn his one-day hack into a more ambitious
project.  Within a few weeks, we had compiled a list of Free software
packages used for teaching in our first, second and third year courses, and
started writing a set of scripts that would build a ``Juppix'' Live-CD
image with no human interaction.

Writing the scripts turned out to be a pointlessly exciting exercice.  At
the time, {\em Debian GNU/Linux}, the Linux distribution on which Knoppix,
and hence Juppix, is based, was not quite ready for unattended
installation.  Some packages refused to install or uninstall without human
interaction, and others had configuration defaults that would lead to
a broken installation.  Hence, our scripts needed to go around patching the
Knoppix filesystem before starting to uninstall and then install software.

\paragraph{Reconfiguring software} Another issue was that of the default
configuration.  Since the home directory of a LiveCD's user is typically
not persistent, we wanted to provide our students with a comfortable
default configuration, one that, as noted above is as similar as possible
to the configuration of the machines in the university's computer rooms.
In particular, we wanted a clean KDE desktop without annoying pop-up
windows\footnote{Nonetheless, the authors bear no grudge towards the author
  of the infamous {\em KTip\/} applet.}, and we wanted the default
shortcuts to follow the well-established pattern of the university
configuration (a terminal, an Emacs editor, a keyboard layout switcher,
etc.).  We also wanted the shipped web browsers to include set of bookmarks
that were likely to be useful, such as pointers to timetables, lecture
notes, exam schedules, and, of course, to {\em xkcd}.

Older software turned out to be easy to reconfigure.  For example,
reconfiguring Emacs was a simple matter of adding a ``\verb|.emacs|'' file
to the default user's home directory ; reconfiguring the shell was done by
appending some code to the default ``\verb|.bashrc|''.

The KDE desktop, however, had clearly not been designed to be reconfigured
from a script.  KDE distributes its configuration files in a complex
hierarchy rooted at a directory ``\verb|.kde|'' in the user's home
directory, and finding the exact location of {\em Konqueror\/}'s bookmarks
or of {\em Kicker\/}'s icons and working out these files' syntax turned out
to be a tedious and error-prone task.  Our scripts remain brittle with
respect to KDE version changes.

The {\em FireFox\/} web browser turned out to be an interesting case.
A {\em FireFox\/} user's configuration is stored in a directory the name of
which is randomly chosen on a per-user basis; the authors have no idea why
this is done, and suspect that the FireFox developers like to be
gratuitiously annoying.  However, once the name {\em du jour\/} of the
configuration directory has been identified, configuring FireFox is
a simple matter of modifying two relatively well documented files
(``\verb|bookmarks.html|'' and ``\verb|prefs.js|'').

\paragraph{Legal issues}

In the small-scale experiments performed in 2004, we had ignored some
licensing issues, notably with the distribution of the Java ``Development
Kit'' (JDK).  Since the 2005 version was to be more widely distributed, we
decided we could not afford an uncertain legal situation.

We considered a number of alternate implementations of Java that didn't
have distribution issues.  Unfortunately, all of the implementations that
we experimented with at the time exhibited differences with Sun's
implementation that, while minor, were significant enough to confuse
beginners.

Sun's legal department were friendly and sympathetic, and encouraged us to
include their Java implementation.  Unfortunately, they appeared to be
unwilling to state anything clearly.  It took the first author two weeks of
consistent harrassment over e-mail across 9 timezones before he managed to
trick a very nice lady into producing a statement that was clear enough for
our Head of Department.

We also had to give up on some software for legal reasons.  At the time,
our third-year course in Logic Programming used a well-known Prolog
implementation from a Portuguese university.  After investigating its
licensing, it turned out that it contained an advertising clause, which we
find obnoxious, and that some of the bundled libraries had a distinct
license that did not allow redistribution at all.  A painful week of e-mail
exchanges did not allow us to converge to a satisfactory solution --- the
authors did not appear to care what we did with their software, but they
were not willing to make any changes to their licensing.  All versions of
Juppix to date have shipped without a Prolog implementation.

\paragraph{Juppix 2005}

By the end of October 2005 (two weeks after the start of the semester), we
had a set of scripts that generated a CD image that we did not fear would
not undermine our professional façade.  We burnt 40 CDs by hand, and
distributed them to our first-year students, 2 per tutorial group, asking
them to share and copy the CDs among themselves.

\section{Fixing mistakes}

In the process of preparing and distributing the 2005 version of Juppix, we
had made a number of fairly serious mistakes, which we felt we needed to
fix for 2006.

\paragraph{Systematic distribution}

While most of our students have a fair amount of experience with copying
CDs, asking them to organise distribution of Juppix turned out to be
a mistake.  In many cases, a tutor managed to make sure that all of hist
students got a copy; in some groups, however, the original CDs ended up
stuck at somebody's home after just a few copies had been made.

Additionally, the fact that our distributon was selective created a feeling
of injustice among the students who weren't able to get a copy.  There was
even a formal complaint to our Head of Department, which he dutifully
dumped on our desks before ignoring it.

At the encouragement of our Head of Department, we decided that if we were
to distribute CDs to our students, we would distribute them to everyone,
and order at least as many pro\-fes\-sional\-ly-pressed CDs as we had
first-year students.

\paragraph{Test suite}

During the development of the 2005 version of Juppix, we lost a lot of time
dealing with silly mistakes.  We would encounter an issue with e.g.\ the
Caml mode not being configured properly in Emacs, which we would promptly
fix, only in order to discover that, although they had been working
perfectly the week before, the locally-installed Java classes were now
broken.

We therefore proposed a fifth year project that consisted in developing
a test suite for Juppix.  The project was taken up by Glenn Rolland, who
wrote a comprehensive set of shell scripts that tested the ``serious''
aspects of Juppix, notably the Emacs setup, the C and Java compilers and
libraries, and the Apache/PHP installation.  The test suite can be run on
a Juppix instance connected to the Internet by simply downloading the
scripts and running a shell script.  Being able to comprehensively test
what we feel is the core of Juppix whenever we build a new image has
significantly reduced our chances of developing a stomach ulcer.

Working with Rolland turned out to be even more helpful than expected.  As
the first outsider who had ever had to run our scripts, Rolland did a fair
amount of work to clean them up and make them easier to use.  He automated
the creation of bookmark files, so that we now could generate bookmarks for
Konqueror and FireFox from a single source.  His project report
\cite{rolland:rapport} remains the best description of the 2006 version of
Juppix.

\section{Results}

In October 2006, we designed cover art for Juppix and ordered a batch of
2000 pro\-fes\-sional\-ly-pressed Juppix CDs; this cost our department
roughly \euro2500.  A few weeks later, we received five boxes of 400 CDs
each.  We set-up an e-mail alias for support (populated with a few Ph.D.\
students with good Linux experience), and gave out 40 CDs to every tutorial
group.

\paragraph{Numbers} Over two years later, three generations of first-year
students have received Juppix CDs, and the 2000 units are mostly gone.
Over the last three academic years, we estimate that roughly 1000 copies
have been distributed to our first year students, an extra 500 copies went
to students in higher years and in other departments (notably
Bio-Informatics), and some 150 copies were distributed in other
universities.  Additionally, some 300 copies have been given out at various
public events, notably the ``{\em Fête de la Science},'' an annual
exposition open to the general public (and their children).

We are aware of at least two instances of other universities burning their
own runs of unmodified Juppix CDs.  On the other hand, we are not aware of
anyone modifying our scripts to produce Juppix-like CDs.

\paragraph{Students} Students tend to consider the use of Juppix as being
a compulsory part of their education, and they are very surprised when they
learn that we don't object to their using a stock Linux distribution
instead.  While we have not performed any large-scale surveys, our tutors'
estimates and the amount of traffic on the support alias indicate that
usage rates are low at the beginning of the semester (in the 10\,\% range),
and increase dramatically in the last week before the November exams.
After they become more proficient with Linux, most students end up
installing a stock Linux distribution, typically in their second or third
year.

An unexpected but pleasant side effect is that the authors have earned the
reputation of being Linux gurus, and hence have acquired the moral right to
be Free Software fanatics.  In past years, the first author would receive
complaints that his Operating Systems lecture was too strongly centered
around Unix, and didn't contain enough ``useful'' information, i.e.\
information that is directly applicable to Microsoft Windows.  He is
pleased to report that his students are now surprised when they discover
that his lecture does mention Windows after all, and that he is not the
Unix monoculturalist that they expected him to be.

\paragraph{Teaching staff} Our teaching colleagues received Juppix in an
overwhelmingly positive manner.  A number of lecturers now systematically
recommend Juppix at the beginning of their lectures, and we regularly
receive requests for small batches of Juppix from tutors whom we don't
personally know.

Some of our colleagues are Unix veterans, and consider Free Unices with
a mixture of sarcasm and bitterness (``You youngsters are just repeating
what we did in the eighties, without giving proper credit'').  They were
clearly impressed with our ability to compile a CD of useful software in
a few months of our copious free time, and distribute it legally to
hundreds of students at a reasonable cost.  The authors are pleased to
report that they are now allowed to speak to the Unix veterans without
first being spoken to\footnote{We have been avoiding the painful subjects,
  notably software patents.}.

\paragraph{Technical staff} We were uncertain of the reactions of the
system administrators, who, we feared, might consider that we were
encroaching on their territory.  Just like the teaching staff, however,
they soon realised that Juppix would reduce their support load, and were
quick to recommend it to their users.  The only complaints we have received
from system administrators were about Juppix not being sufficiently
up-to-date.

\section{Current work: Juppix 2009}

Over the last years, the 2006 version of Juppix has become increasingly out
of date.  While the software it contains is still useful, it fails to boot
on an increasing number of recent machines, which is a serious issue as
first year students tend to have very new hardware.  Additionally, a number
of hardware and software changes have occurred in the intervening years,
which makes it possible to build a better Juppix now than it was in 2006.

\paragraph{New storage media} The price of DVD duplication has dropped
significantly, to the extent that it becomes reasonable to consider
distributing a live-DVD rather than a live-CD.  While we have no intention
of significantly expanding the functionality of Juppix, the increase in
capacity (4.7\,GB instead of 750\,MB) will simplify our work, as we will no
longer have to compromise between the conflicting goals of providing
a complete programming environment and keeping the image size below
750\,MB.

Additionally, most of our students own flash keys of 1\,GB or more, rather
than the 128\,MB that were typical in 2006.  It becomes useful to offer
a version of Juppix that can be booted off a flash key, thus avoiding the
slow speed of optical media.

\paragraph{Other changes}

Since Wifi access points are more common now than they were in 2006, we
hope to provide good support for common wireless hardware.  This will
probably require shipping with a Linux kernel different from Debian's
default.

The licensing conditions of Sun's Java implementation have changed.  The
stock 1.6 JDK is still proprietary software, but can be redistributed under
fairly reasonable conditions, and there exists a version called {\em
  OpenJDK\/} that is Free Software.  We are currently evaluating the use of
{\em OpenJDK}, with the option of switching back to the JDK if it proves
unsuitable.

\paragraph{Live-Helper} Since we compiled Juppix 2006, Live-CD technology
has improved significantly.  Of particular note is the inclusion in the
Debian GNU/Linux distribution of the {\em live-helper\/} and {\em
  live-initramfs\/} packages, which allow the reasonably automated creation
of a live-CD, a live-DVD or a live-USB key.

We are currently working on a version of Juppix based on the {\em
  live-helper\/} scripts and on Debian's {\em lenny\/} distribution.  The
new version will still be built by a custom set of scripts, but these
scripts will be driving the {\em live-helper\/} scripts rather than
modifying an existing image.  We expect to physically distribute a Live-DVD
version to our students; for people wishing to download an image, we will
offer at least two versions:
\begin{itemize}
\item an {\em iso-large\/} version, suitable for burning on DVD media, and
  including full documentation, graphical development environment,
  a complete TeX system and a word processor;
\item an {\em iso-small\/} version, suitable for burning on CD media,
  including reasonable documentation, Emacs, and a minimal TeX system.
\end{itemize}
At the time of writing, it is not clear to us whether we will need to ship
separate images for flash keys, or whether we will be able to install on
a flash key from a live system booted from DVD.

\section{Conclusions}

Four years' experience with Juppix has taught us some important lessons.
\begin{itemize}
\item If your underground project is cool, tell your boss about it.  If you
  think it's cool, he'll likely think so too.
\item If you distribute something to your students, either distribute it to
  all of them, or don't do it at all.  Students are quick to perceive
  differences in treatment as injustice.
\item Don't lose time arguing with people who think that advertising
  clauses are a good idea, or who forbid redistribution.  Dropping their
  software is easier on the nerves.
\item Students are cheap, but the private sector is cheaper.  Don't burn
  your own CDs, get them pro\-fes\-sional\-ly pressed.
\item Don't be afraid of the old Unix crowd.
\item Don't be afraid of your system administrator.
\end{itemize}

\section*{Software availability}

Ready-to-burn Juppix image files, the scripts used to build Juppix, and the
test suite are available from
\url{http://www.pps.jussieu.fr/~jch/software/juppix}.


\begin{thebibliography}{}
\frenchspacing

\bibitem{demolinux} Vincent Balat, Roberto DiCosmo and Jean-Vincent
  Loddo.  {\em The DemoLinux manual}.  September 1999.

\bibitem{debian-live} The Debian Live Project.  {\em Debian Live
    Manual}.  2008.

\bibitem{knoppix} Klaus Knopper.  {\em Knoppix 3.6}.  August 2004.

\bibitem{yggdrasil-live-cd} Adam J. Richter.  {\em ANNOUNCE: Alpha
  release of turnkey Linux/GNU/X system on CDROM}.  Linux-Activists
  mailing list.  24 November 1992.

\bibitem{rolland:rapport} Glenn Rolland. {\em Juppix}.  Rapport de
  projet long.  Master 2 d'Ingénierie Informatique de l'Université de
  Paris VII -- Denis Diderot.  2006.  Available online from
  \url{http://www.pps.jussieu.fr/~jch/software/juppix/}.  (In French.)

\bibitem{ubuntu-live-cd} Ubuntu Documentation Project.  {\em Ubuntu
    Installation Guide.}  Version 6.06.  June 2006.

\bibitem{classe-deug} Jean-Baptiste Yunés.  {\em La classe Deug}.
  2002.  Available online from
  \url{http://www.liafa.jussieu.fr/~yunes/deug/Deug/}.

\nonfrenchspacing
\end{thebibliography}
\end{document}